\documentclass{fundam}

\publyear{25}
\papernumber{6}
\volume{195}
\issue{1-4}
\theDOI{10.46298/fi.12347}

\versionForARXIV

\usepackage[numbers,sort&compress]{natbib}
\usepackage{dsfont,hyperref}

\newcommand{\AB}{A}
\newcommand{\bigjoin}{\bigsqcup}
\newcommand{\cnv}[1]{{#1}^{\sf T}}
\newcommand{\cnvop}{\cnv{}}
\newcommand{\cl}[1]{\cpl{\cpl{#1}}}
\newcommand{\clop}{\cl{\rule{0em}{1.2ex}~\,}}
\newcommand{\cpl}[1]{\overline{#1}}
\newcommand{\cplop}{\cpl{\rule{0em}{1.2ex}~\,}}
\renewcommand{\iff}{\Leftrightarrow}
\renewcommand{\implies}{\Rightarrow}
\newcommand{\I}{\mathrm{I}}
\newcommand{\ind}[3]{{#1}_{{#2},{#3}}}
\newcommand{\IP}{\mathrm{IP}}
\newcommand{\join}{\sqcup}
\newcommand{\lbot}{\bot}
\newcommand{\lleq}{\sqsubseteq}
\newcommand{\lone}{1}
\newcommand{\ltop}{\top}
\newcommand{\M}{\mathrm{M}}
\newcommand{\mat}[4]{\binom{#1\,#2}{#3\,#4}}
\newcommand{\meet}{\sqcap}
\newcommand{\nAB}{C}
\newcommand{\NN}{\mathds{N}}
\newcommand{\ZZ}{\mathds{Z}}

\newtheorem{counterexample}{Counterexample}
\newtheorem{thm}{Theorem}
\newenvironment{prf}{\trivlist\PRstyle\item[]{\bfseries Proof:}}{\endtrivlist}

\newcommand{\mailto}[1]{\href{mailto:#1}{#1}}

\begin{document}

\title{Cardinality and Representation of Stone Relation Algebras}
\author{Hitoshi Furusawa \\
        Department of Science \\
        Kagoshima University \\
        Kagoshima, Japan \\
        \mailto{furusawa@sci.kagoshima-u.ac.jp}
        \and
        Walter Guttmann \\
        Department of Computer Science and Software Engineering \\
        University of Canterbury \\
        Christchurch, New Zealand \\
        \mailto{walter.guttmann@canterbury.ac.nz}}
\runninghead{H. Furusawa, W. Guttmann}{Cardinality and Representation of Stone Relation Algebras}
\maketitle

\vspace*{-5ex}

\begin{abstract}
  Previous work has axiomatised the cardinality operation in relation algebras, which counts the number of edges of an unweighted graph.
  We generalise the cardinality axioms to Stone relation algebras, which model weighted graphs, and study the relationships between various axioms for cardinality.
  This results in simpler cardinality axioms also for relation algebras.
  We give sufficient conditions for the representability of Stone relation algebras and for Stone relation algebras to be relation algebras.
\end{abstract}


\urlstyle{rm}

\section{Introduction}

Relation algebras have been introduced by Tarski to algebraically describe properties of binary relations and capture a fragment of first-order logic \cite{Tarski1941}.
Since binary relations are closely connected to graphs, relation algebras have been used for studying graphs and graph algorithms \cite{SchmidtStroehlein1989}.
For this purpose, counting the number of vertices and edges is important and achieved by extending algebras/categories of relations with a cardinality operation and suitable axioms \cite{Kawahara2006,BerghammerDanilenkoHoefnerStucke2016}.
Stone relation algebras have been introduced in \cite{Guttmann2017b} to model weighted graphs; they generalise relation algebras which capture only unweighted graphs.

This paper combines and continues the above lines of research by investigating how the cardinality operation generalises from relation algebras to Stone relation algebras.
It also studies the representation of (Stone) relation algebras extended with a cardinality operation.
A relation algebra is representable if it is isomorphic to an algebra of binary relations with the usual relational operations.
Several sufficient conditions for representability are known \cite{JonssonTarski1952,Tarski1953,MadduxTarski1976,Maddux1978,SchmidtStroehlein1985,TarskiGivant1987,Maddux1991}, but it is open whether relation algebras with a cardinality operation are representable.
Section \ref{section.relation-algebras} contributes both positive and negative answers to this question.
The paper also gives a sufficient condition for the representability of Stone relation algebras.

Cardinality and representability are treated together in this paper because both notions are linked through the concept of atoms.
In particular, the paper investigates a cardinality operation that counts the number of atoms below each element.
Requirements for the cardinality operation then affect properties of atoms; for example, see Theorem \ref{theorem.card-sra-ra}.
On the other hand, it is known that certain postulates for atoms provide sufficient conditions for a relation algebra to be representable.
It is therefore a natural question whether requirements for the cardinality operation make a relation algebra representable.

The main results of this paper are:
\begin{itemize}
\item Every Stone relation algebra satisfying an additional axiom can be represented by lattice-valued matrices (Theorem \ref{theorem.representation}).
      This is in line with previously known representations of Dedekind categories and relation algebras.
\item Axioms of the cardinality operation for relation algebras can be used in Stone relation algebras, but some of them need to be adjusted and different combinations of new axioms can be considered (Figure \ref{figure.cardinality-axioms}).
\item The operation that counts the number of atoms below an element satisfies most cardinality axioms in atomic Stone relation algebras with finitely many atoms (Theorem \ref{theorem.nAB-card}).
      More generally, this paper studies five conditions and identifies which of them are sufficient for each of the cardinality axioms (Theorem \ref{theorem.nAB-card-sufficient} and Counterexample \ref{counterexample.nAB-card-sufficient}).
\item Every simple and atomic Stone relation algebra with finitely many atoms and an additional condition is a relation algebra (Theorem \ref{theorem.atom-sra-ra}).
      This unexpected result affects the generalisation of cardinality operations to Stone relation algebras.
\item Every simple Stone relation algebra with finitely many atoms and a cardinality operation that counts the number of atoms below an element is a relation algebra (Theorem \ref{theorem.card-sra-ra}).
      This, too, affects the generalisation of cardinality to Stone relation algebras.
\item There are equivalent, simpler formulations of two of the cardinality axioms in relation algebras (Theorems \ref{theorem.card-equivalent-sra} and \ref{theorem.card-equivalent-ra}).
      This considerably expands the search space for suitable axioms for Stone relation algebras, where these formulations are no longer equivalent.
\item In an atomic relation algebra with finitely many atoms, any operation satisfying the cardinality axioms indeed counts the number of atoms below an element (Theorem \ref{theorem.card-nAB}).
      Hence counting the number of atoms is a canonical form of cardinality operations in relation algebras.
\item Every simple and atomic relation algebra with finitely many atoms and a cardinality operation is representable (Theorem \ref{theorem.card-ra}).
\item There is an atomic relation algebra with finitely many atoms and a cardinality operation that does not satisfy typical sufficient conditions for representability (Counterexample \ref{counterexample.representable}).
      Still, this counterexample is representable.
\end{itemize}
All theorems of this paper have been formally verified in Isabelle/HOL \cite{NipkowPaulsonWenzel2002}; see \cite{Guttmann2023} for these proofs.
The Isabelle/HOL theories also contain most of the counterexamples discussed in this paper.
The theories complement existing theories based on Stone relation algebras, which provide models of weighted and unweighted graphs and correctness guarantees for graph algorithms.

\section{Basic Definitions and Properties}

In this section we define Stone relation algebras \cite{Guttmann2017b} and give basic properties.
For further details about relation algebras see, for example, \cite{Tarski1941,SchmidtStroehlein1989,HirschHodkinson2002,Maddux2006}.

A \emph{bounded semilattice} $(S,\join,\lbot)$ is a set $S$ with an associative, commutative and idempotent binary operation $\join$ and a constant $\lbot$ that is a unit of $\join$.
We write $\bigjoin P$ for applying $\join$ to the elements of a finite non-empty set $P \subseteq S$.
A \emph{bounded lattice} $(S,\join,\meet,\lbot,\ltop)$ comprises bounded semilattices $(S,\join,\lbot)$ and $(S,\meet,\ltop)$ such that the absorption laws $x \join (x \meet y) = x = x \meet (x \join y)$ hold for all $x, y \in S$.
A \emph{bounded distributive lattice} is a bounded lattice $S$ such that the distributivity law $x \join (y \meet z) = (x \join y) \meet (x \join z)$ holds for all $x, y, z \in S$.
Distributivity of $\meet$ over $\join$ follows from this.
The lattice order is defined by $x \lleq y \iff x \join y = y$ for all $x, y \in S$.
A (distributive) \emph{p-algebra} $(S,\join,\meet,\cplop,\lbot,\ltop)$ is a bounded (distributive) lattice $(S,\join,\meet,\lbot,\ltop)$ with a unary pseudocomplement operation $\cplop$ that satisfies $x \meet y = \lbot \iff x \lleq \cpl{y}$ for all $x, y \in S$.
A \emph{Stone algebra} is a distributive p-algebra $S$ such that $\cpl{x} \join \cl{x} = \ltop$ for all $x \in S$.

A \emph{monoid} $(S,\cdot,\lone)$ is a set $S$ with an associative binary operation $\cdot$ and a constant $\lone$ that is a left and right unit of $\cdot$.
We abbreviate $x \cdot y$ as $x y$.
An \emph{idempotent semiring} $(S,\join,\cdot,\lbot,\lone)$ comprises a bounded semilattice $(S,\join,\lbot)$ and a monoid $(S,\cdot,\lone)$ such that $\cdot$ distributes over $\join$ and $\lbot$ is a zero of $\cdot$.
An \emph{idempotent semiring with involution} $(S,\join,\cdot,\cnvop,\lbot,\lone)$ is an idempotent semiring $(S,\join,\cdot,\lbot,\lone)$ with a unary involution $\cnvop$ that satisfies $\cnv{x}\cnvop = x$ and $\cnv{(x y)} = \cnv{y} \cnv{x}$ and $\cnv{(x \join y)} = \cnv{x} \join \cnv{y}$ for all $x, y \in S$.
A \emph{Stone relation algebra} $(S,\join,\meet,\cdot,\cplop,\cnvop,\lbot,\ltop,\lone)$ comprises a Stone algebra $(S,\join,\meet,\cplop,\lbot,\ltop)$ and an idempotent semiring with involution $(S,\join,\cdot,\cnvop,\lbot,\lone)$ such that $\cl{\lone} = \lone$ and $\cl{x y} = \cl{x} \, \cl{y}$ and $x y \meet z \lleq x (y \meet \cnv{x} z)$ for all $x, y, z \in S$.
A \emph{relation algebra} is a Stone relation algebra $S$ such that $\cl{x} = x$ for all $x \in S$.

Let $S$ be a Stone relation algebra and let $x \in S$.
We call $x$
\begin{itemize}
\item \emph{univalent} if $\cnv{x} x \lleq \lone$;
\item \emph{total} if $\lone \lleq x \cnv{x}$;
\item \emph{a mapping} if $x$ is univalent and total;
\item \emph{injective} if $x \cnv{x} \lleq \lone$;
\item \emph{surjective} if $\lone \lleq \cnv{x} x$;
\item \emph{bijective} if $x$ is injective and surjective;
\item a \emph{vector} if $x \ltop = x$;
\item a \emph{covector} if $\ltop x = x$;
\item a \emph{point} if $x$ is a bijective vector;
\item a \emph{rectangle} if $x \ltop x \lleq x$;
\item \emph{simple} if $\ltop x \ltop = \ltop$;
\item an \emph{atom} if $x \neq \lbot$ and $y = x$ for each $y \in S$ such that $\lbot \neq y \lleq x$.
\end{itemize}
Stone relation algebra $S$ is \emph{simple} if every element in $S$ except $\lbot$ is simple.
For relation algebras this condition is equivalent to being simple in the universal-algebraic sense \cite{JonssonTarski1952}, which is why we apply the term `simple' also to elements.
The full algebra of relations over the two-element base set $\{ a, b \}$ contains two points $\{ (a,a), (a,b) \}$ and $\{ (b,a), (b,b) \}$ and four atoms $\{ (a,a) \}$, $\{ (a,b) \}$, $\{ (b,a) \}$ and $\{ (b,b) \}$.

The following result summarises basic properties, which are useful to establish the results in this paper.
\begin{thm}
  \label{theorem.basic}
  Let $S$ be a Stone relation algebra and let $a, b, p, x, y, z \in S$.
  Then
  \begin{enumerate}
  \item $x \meet y = \lbot \iff \cnv{x} y = \lbot$ if $x$ is a vector.
  \item $\cnv{x} x = \ltop$ if $x$ is a surjective vector.
  \item $\ltop = \bigjoin P \iff \lone = \bigjoin \{ p \cnv{p} \mid p \in P \}$ if $P \subseteq S$ is a non-empty finite set of points.
  \item $p \cnv{p} = p \meet \lone$ if $p$ is a point.
  \item $x = p \cnv{p} x$ if $p$ is a point and $x \lleq p$.
  \item $\lone \meet x \ltop = \lone \meet x \cnv{x} = \lone \meet \ltop \cnv{x}$.
  \item $x y \meet z = (x \meet z \cnv{y}) y$ if $y$ is univalent.
  \item $x y \meet z = x (y \meet \cnv{x} z)$ if $x$ is injective.
  \item $a$ is an atom if and only if $\cnv{a}$ is an atom.
  \item $a \ltop \meet \lone$ and $\ltop a \meet \lone$ are atoms if $a$ is an atom.
  \item $a \meet b = \lbot$ if $a$ and $b$ are different atoms.
  \item Exactly one of $a \lleq x$ and $a \lleq \cpl{x}$ holds if $a$ is an atom.
  \item $a \lleq x \join y \iff a \lleq x \vee a \lleq y$ if $a$ is an atom.
  \item $a \lleq x \join y$ if and only if exactly one of $a \lleq x$ and $a \lleq y$ holds, if $a$ is an atom and $x \meet y = \lbot$.
  \end{enumerate}
\end{thm}

\begin{counterexample}
  \label{counterexample.atom_univalent}
  There are a relation algebra $S$ and $x \in S$ such that the following statement does not hold: $x$ is univalent if $x$ is an atom.
\end{counterexample}
\begin{shortproof}
  Nitpick \cite{BlanchetteNipkow2010} found the following counterexample.
  The set $\{\lbot,\lone,\cpl{\lone},\ltop\}$ of relations over a three-element base set forms a relation algebra which is a subalgebra of the full algebra of relations.
  In this subalgebra, $\cpl{\lone}$ is an atom but not univalent.
\end{shortproof}

\section{Representability of Stone Relation Algebras}
\label{section.representation}

In this section we give a sufficient condition for a Stone relation algebra $S$ to be representable.

For a lattice $L$, an $L$-valued binary relation on base set $A$ is a function from $A^2$ to $L$, or equivalently an $A \times A$ matrix with entries from $L$.
The usual binary relations arise as the special case where $L$ is the two-element Boolean algebra.
A Stone relation algebra is representable if it is isomorphic to a Stone relation algebra of Stone-algebra-valued binary relations with the usual relational operations for lattice-valued binary relations.

Several sufficient conditions for the representability of relation algebras are known \cite{JonssonTarski1952,Tarski1953,MadduxTarski1976,Maddux1978,SchmidtStroehlein1985,TarskiGivant1987,Maddux1991}.
One of these conditions is called point axiom \cite{SchmidtStroehlein1985}.
Modifying the concept of points and the point axiom, representation theorems for Dedekind categories can be shown \cite{Furusawa1998,KawaharaFurusawaMori1999}.
We take a similar approach here.

Element $x \in S$ is an \emph{ideal} if $x$ is a vector and a covector.
Let $\I(S)$ be the set of ideals in $S$.
The following result gives basic properties of ideals.
\begin{thm}
  Let $S$ be a Stone relation algebra.
  Let $x, y, z \in S$.
  Then
  \begin{enumerate}
  \item $x$ is an ideal if and only if $\ltop x \ltop = x$.
  \item $\cnv{x} y z$ is an ideal if $x$ and $z$ are vectors.
  \item Ideals are closed under $\cdot$, $\join$, $\meet$, $\cnvop$ and $\cplop$.
  \item $\lbot$ and $\ltop$ are ideals.
  \item $x y = x \meet y$ if $x$ and $y$ are ideals.
  \item $\I(S)$ is a Stone relation algebra where $\cdot$ is $\meet$ and $\lone$ is $\ltop$.
  \item $x = \cnv{x}$ if $x$ is an ideal.
  \item $y z \meet x = (y \meet x) (z \meet x)$ if $x$ is an ideal.
  \end{enumerate}
\end{thm}

Element $p \in S$ is an \emph{ideal-point} if $p$ is a point and $q x \lleq p$ implies $q \lleq p$ for all points $q \in S$ and ideals $x \in S$ with $x \neq \lbot$.
Let $\IP(S)$ be the set of ideal-points in $S$.
The following result characterises different ideal-points.
\begin{thm}
  Let $S$ be a Stone relation algebra.
  Then
  \begin{enumerate}
  \item $p \meet q = \cnv{p} q = \lbot$ for each $p, q \in \IP(S)$ such that $p \neq q$.
  \item Every point is an ideal-point if $S$ is simple.
  \item $p \in S$ is an ideal-point if and only if $p$ is a point and $q \lleq p$ or $q \lleq \cpl{p}$ for all points $q \in S$.
  \end{enumerate}
\end{thm}

\begin{counterexample}
  \label{counterexample.non-simple-ideal-point}
  ~
  \begin{enumerate}
  \item There is a non-simple Stone relation algebra $S$ with $\ltop \neq \lone$ and an ideal-point $p \in S$.
  \item There is a non-simple relation algebra with points but no ideal-points.
  \end{enumerate}
\end{counterexample}

\begin{shortproof}
  \begin{enumerate}
  \item The set $S = \{0,\frac{1}{2},1\}$ of numbers with their natural order forms a Stone relation algebra where $\cnvop$ is identity, $\cdot$ is $\meet$, $\cpl{0} = 1$ and $\cpl{\frac{1}{2}} = \cpl{1} = 0$.
        The $2 \times 2$ matrices with entries from $S$ form a Stone relation algebra, where $\ltop = \mat{1}{1}{1}{1}$, $\lone = \mat{1}{0}{0}{1}$ and $\cdot$ is matrix product using $\max$ and $\min$ instead of addition and multiplication \cite{Guttmann2017b}.
        In this matrix algebra, $\mat{1}{1}{0}{0}$ and $\mat{0}{0}{1}{1}$ are ideal-points, but only the matrices which contain a $1$ entry are simple.
  \item Let $R$ be the full algebra of relations over the two-element base set $\{ a, b \}$.
        Then $R$ has two points $p = \{ (a,a), (a,b) \}$ and $q = \{ (b,a), (b,b) \}$.
        The Cartesian product $R \times R$ is a relation algebra where operations are applied componentwise.
        This algebra is not simple and has four points $(p,p)$, $(p,q)$, $(q,p)$ and $(q,q)$ but no ideal-points.
  \end{enumerate}
\end{shortproof}
Stone relation algebra $S$ satisfies the \emph{point axiom} if $\IP(S)$ is finite and non-empty and $\ltop = \bigjoin \IP(S)$.
The following result gives consequences of the point axiom.
\begin{thm}
  \label{theorem.point-axiom}
  Let $S$ be a Stone relation algebra satisfying the point axiom.
  Let $x \in S$.
  Then
  \begin{enumerate}
  \item $\lone = \bigjoin \{ p \cnv{p} \mid p \in \IP(S) \}$.
  \item $x = \bigjoin \{ p \cnv{p} x q \cnv{q} \mid p, q \in \IP(S) \}$.
  \item For each atom $a \in S$ there is a $p \in \IP(S)$ such that $a \lleq p$.
  \item Every point is an ideal-point.
  \end{enumerate}
\end{thm}

Elements of $S$ can be represented by square matrices of type $\M(S) = \I(S)^{\IP(S) \times \IP(S)}$ whose index set is $\IP(S)$ and whose entries are from $\I(S)$.

\eject

\begin{thm}
  \label{theorem.representation}
  Let $S$ be a Stone relation algebra satisfying the point axiom.
  Then
  \begin{enumerate}
  \item $\M(S)$ is a Stone relation algebra where $\join$, $\meet$, $\cplop$, $\lbot$ and $\ltop$ are lifted componentwise and $\cdot$, $\cnvop$ and $\lone$ are
        \begin{align*}
          \ind{(XY)}{p}{q} & = \bigjoin \{ \ind{X}{p}{r} \meet \ind{Y}{r}{q} \mid r \in \IP(S) \} \\
          \ind{\cnv{X}}{p}{q} & = \ind{X}{q}{p} \\
          \ind{\lone}{p}{q} & = \left\{ \begin{array}{ll} \ltop & \mbox{if $p = q$} \\ \lbot & \mbox{if $p \neq q$} \end{array} \right.
        \end{align*}
  \item The functions
        \begin{align*}
          f & : S \to \M(S) & g & : \M(S) \to S \\
          \ind{f(x)}{p}{q} & = \cnv{p} x q & g(X) & = \bigjoin \{ p \ind{X}{p}{q} \cnv{q} \mid p, q \in \IP(S) \}
        \end{align*}
        are isomorphisms between Stone relation algebras $S$ and $\M(S)$.
  \end{enumerate}
\end{thm}

\section{Cardinality in Stone Relation Algebras}
\label{section.cardinality}

The cardinality of a relation is the usual set cardinality, that is, the number of pairs in the relation.
More abstractly, axioms for a cardinality operation have been proposed for Dedekind categories and heterogeneous relation algebras \cite{Kawahara2006,BerghammerDanilenkoHoefnerStucke2016}.
In this section, we study different ways to adapt these axioms to Stone relation algebras.

In a Stone relation algebra $S$, the cardinality operation $\# : S \to \NN \cup \{ \infty \}$ maps each element of $S$ to a natural number or $\infty$.
Extending previous works, we allow relations with infinite cardinalities; we do not distinguish between different types of infinity.
We consider the formulas in Figure \ref{figure.cardinality-axioms} as potential axioms.

We informally discuss the meaning of the candidate axioms.
Axioms (\ref{card_bot}) and (\ref{card_bot_iff}) express that the least element $\lbot$ has cardinality $0$, which describes the lower end of the range of cardinalities.
Axioms (\ref{card_atom}) and (\ref{card_atom_iff}) express that atoms have cardinality $1$, which normalises cardinalities.
Axiom (\ref{card_conv}) expresses that $\cnvop$ does not affect cardinalities.
Axioms (\ref{card_add}) and (\ref{card_iso}) express how cardinality preserves the lattice structure.
Axioms (\ref{card_univ_comp_meet})--(\ref{card_univ_meet_conv}) express inequalities for cardinalities related to the modular law $x \meet y z \lleq y (z \meet \cnv{y} x)$; the univalence assumption yields strong consequences such as Theorems \ref{theorem.card-equivalent-sra}.\ref{theorem.card-equivalent-sra.card_univ_comp_mapping} and \ref{theorem.card-equivalent-sra}.\ref{theorem.card-equivalent-sra.card_point_one}.
Axioms (\ref{card_domain_sym}) and (\ref{card_domain_sym_conv}) specialise (\ref{card_univ_meet_comp}) to the (co)domain of an element represented as an element below $\lone$, which reduces the number of variables and avoids the univalence assumption.
Axioms (\ref{card_top_iff_eq}) and (\ref{card_top_iff_leq}) express that only the greatest element $\ltop$ has maximal cardinality, which describes the upper end of the range of cardinalities.
Axiom (\ref{card_top}) relates cardinalities with pairs of elements below $\lone$, which is typical for representable relation algebras.
Axiom (\ref{card_top_finite}) expresses that $\ltop$ has finite cardinality; omitting it allows us to consider relations with infinite cardinality.

\begin{figure}[p]
  \begin{align}
    & \# \lbot = 0 \tag{\mbox{C1a}} \label{card_bot} \\
    & \forall x \in S : \# x = 0 \iff x = \lbot \tag{\mbox{C1b}} \label{card_bot_iff} \\[2ex]
    & \forall x \in S : \mbox{atom $x$} \implies \# x = 1 \tag{\mbox{C2a}} \label{card_atom} \\
    & \forall x \in S : \mbox{atom $x$} \iff \# x = 1 \tag{\mbox{C2b}} \label{card_atom_iff} \\[2ex]
    & \forall x \in S : \# (\cnv{x}) = \# x \tag{\mbox{C3}} \label{card_conv} \\[2ex]
    & \forall x, y \in S : \# x + \# y = \# (x \join y) + \# (x \meet y) \tag{\mbox{C4a}} \label{card_add} \\
    & \forall x, y \in S : x \lleq y \implies \# x \leq \# y \tag{\mbox{C4b}} \label{card_iso} \\[2ex]
    & \forall x, y, z \in S : \mbox{univalent $x$} \implies \# (\cnv{x} y \meet z) \leq \# (x z \meet y) \tag{\mbox{C5a}} \label{card_univ_comp_meet} \\
    & \forall x, y, z \in S : \mbox{univalent $x$} \implies \# (x \meet y \cnv{z}) \leq \# (x z \meet y) \tag{\mbox{C5b}} \label{card_univ_meet_comp} \\
    & \forall x, y \in S : \mbox{univalent $x$} \implies \# (y x) \leq \# y \tag{\mbox{C5c}} \label{card_comp_univ} \\
    & \forall x, y \in S : \mbox{univalent $x$} \implies \# (x \meet y \ltop) \leq \# y \tag{\mbox{C5d}} \label{card_univ_meet_vector} \\
    & \forall x, y \in S : \mbox{univalent $x$} \implies \# (x \meet y \cnv{y}) \leq \# y \tag{\mbox{C5e}} \label{card_univ_meet_conv} \\[2ex]
    & \forall x \in S : \# (\lone \meet x \cnv{x}) \leq \# x \tag{\mbox{C6a}} \label{card_domain_sym} \\
    & \forall x \in S : \# (\lone \meet \cnv{x} x) \leq \# x \tag{\mbox{C6b}} \label{card_domain_sym_conv} \\[2ex]
    & \forall x \in S : \# x = \# \ltop \iff x = \ltop \tag{\mbox{C7a}} \label{card_top_iff_eq} \\
    & \forall x \in S : \# \ltop \leq \# x \iff x = \ltop \tag{\mbox{C7b}} \label{card_top_iff_leq} \\[2ex]
    & \# \ltop = (\# \lone)^2 \tag{\mbox{C8}} \label{card_top} \\[2ex]
    & \# \ltop \neq \infty \tag{\mbox{C9}} \label{card_top_finite}
  \end{align}
  \caption{Axioms for the cardinality operation}
  \label{figure.cardinality-axioms}
\end{figure}

In \cite{Kawahara2006}, Kawahara has proposed to use the following cardinality axioms in Dedekind categories: (\ref{card_bot_iff}), (\ref{card_conv}), (\ref{card_univ_comp_meet}), (\ref{card_univ_meet_comp}), a version of (\ref{card_add}) formulated as
\begin{align*}
  \forall x, y \in S : \# (x \join y) = \# x + \# y - \# (x \meet y)
\end{align*}
and $\# \lone_I = 1$ where $I$ is the singleton set.
We use (\ref{card_add}) to avoid subtraction.
We use (\ref{card_atom}) instead of $\# \lone_I = 1$ because we are working with algebras rather than categories and so we cannot refer to specific objects.

Examples of the use of the cardinality axioms for reasoning about algebras are given in Section \ref{section.relation-algebras}.
Examples using the cardinality axioms for reasoning in Dedekind categories and in relation algebras are given in \cite{Kawahara2006,BerghammerDanilenkoHoefnerStucke2016}.

\subsection{Atoms Below an Element}
\label{section.atoms-below}

Let $S$ be a Stone relation algebra.
Let $\AB : S \to 2^S$ map each element $x \in S$ to the set of atoms below $x$.
Let $\nAB : S \to \NN \cup \{ \infty \}$ count the number of atoms below each element.
The following result gives basic properties of $\AB$.

\begin{thm}
  Let $S$ be a Stone relation algebra.
  Let $x, y \in S$.
  Then
  \begin{enumerate}
  \item $\AB(x) \subseteq \AB(y)$ if $x \lleq y$.
  \item $\AB(\lbot) = \emptyset$.
  \item $\AB(x \meet y) = \AB(x) \cap \AB(y)$.
  \item $\AB(x) \cap \AB(y) = \emptyset$ if $x \meet y = \lbot$.
  \item $\AB(x \join y) = \AB(x) \cup \AB(y)$.
  \item $\AB(x) = \AB(x \meet y) \cup \AB(x \meet \cpl{y})$ and $\AB(x \meet y) \cap \AB(x \meet \cpl{y}) = \emptyset$.
  \item $\AB(\cl{x}) = \AB(x)$.
  \end{enumerate}
\end{thm}

Stone relation algebra $S$ is \emph{atomic} if every element except $\lbot$ is above some atom, that is, for each $x \in S$ such that $x \neq \lbot$ there is an atom $a \in S$ such that $a \lleq x$.
In other words, $\AB(x) \neq \emptyset$ for each $x \neq \lbot$.

\begin{thm}
  \label{theorem.nAB-card}
  Let $S$ be a Stone relation algebra.
  Then $\nAB$ satisfies axioms (\ref{card_bot}), (\ref{card_atom}) and (\ref{card_conv})--(\ref{card_iso}).
  If $S$ is atomic, $\nAB$ also satisfies axioms (\ref{card_bot_iff}) and (\ref{card_univ_comp_meet})--(\ref{card_domain_sym_conv}).
\end{thm}

Hence $\nAB$ satisfies (appropriate generalisations of) Kawahara's cardinality axioms in atomic Stone relation algebras.

\subsection{Relationships Between Cardinality Axioms}

Figure \ref{figure.cardinality-axioms} lists a number of alternative axioms.
The following result states connections between the axioms and consequences in Stone relation algebras.

\begin{thm}
  \label{theorem.card-equivalent-sra}
  Let $S$ be a Stone relation algebra.
  Then
  \begin{enumerate}
  \item $(\ref{card_bot_iff}) \implies (\ref{card_bot})$.
  \item $(\ref{card_conv}) \wedge (\ref{card_comp_univ}) \implies (\ref{card_univ_comp_meet})$.
  \item $(\ref{card_univ_meet_comp}) \vee (\ref{card_univ_meet_conv}) \implies (\ref{card_domain_sym})$.
  \item Assume (\ref{card_conv}).
        Then $(\ref{card_domain_sym}) \iff (\ref{card_domain_sym_conv})$.
  \item Assume (\ref{card_conv}) and (\ref{card_iso}).
        Then $(\ref{card_univ_comp_meet}) \iff (\ref{card_comp_univ})$.
  \item Assume (\ref{card_iso}).
        Then $(\ref{card_univ_meet_comp}) \iff (\ref{card_univ_meet_vector}) \implies (\ref{card_univ_meet_conv})$.
  \item Assume (\ref{card_iso}) and (\ref{card_comp_univ}).
        Then $(\ref{card_univ_meet_comp}) \iff (\ref{card_univ_meet_vector}) \iff (\ref{card_univ_meet_conv}) \iff (\ref{card_domain_sym})$.
  \item Assume (\ref{card_iso}).
        Then $\# x \leq \# \ltop$ for each $x \in S$.
        In particular, $(\ref{card_top_iff_eq}) \iff (\ref{card_top_iff_leq})$.
  \end{enumerate}
  Let $x, y \in S$.
  Then
  \begin{enumerate}\setcounter{enumi}{8}
  \item Assume (\ref{card_univ_meet_comp}) and (\ref{card_comp_univ}).
        Then $\# (x y) = \# x$ if $x$ is univalent and $y$ is a mapping.
        \label{theorem.card-equivalent-sra.card_univ_comp_mapping}
  \item Assume (\ref{card_conv}) and (\ref{card_univ_meet_comp}) and (\ref{card_comp_univ}).
        Then $\# x = \# \lone$ if $x$ is a point.
        \label{theorem.card-equivalent-sra.card_point_one}
  \end{enumerate}
  Assume (\ref{card_bot}) and (\ref{card_add}).
  Let $X \subseteq S$ be finite and non-empty.
  Then
  \begin{enumerate}\setcounter{enumi}{10}
  \item $\# (x \join y) = \# x + \# y$ if $x \meet y = \lbot$.
  \item $\# \bigjoin X = \sum_{x \in X} \# x$ if $x \meet y = \lbot$ for each $x, y \in X$ such that $x \neq y$.
  \item $\# \bigjoin X = \sum_{x \in X} \# x$ if each $x \in X$ is an atom.
        \label{theorem.card-equivalent-sra.card_dist_sup_atoms}
  \end{enumerate}
\end{thm}

The following result states connections that hold in relation algebras.
Notably, (\ref{card_iso}) follows from Kawahara's axioms in relation algebras, but needs to be axiomatised separately in Stone relation algebras.
With this axiom, all properties in the range (\ref{card_comp_univ})--(\ref{card_domain_sym_conv}) follow by the previous result.

\begin{thm}
  \label{theorem.card-equivalent-ra}
  Let $S$ be a relation algebra.
  Then
  \begin{enumerate}
  \item $(\ref{card_bot}) \wedge (\ref{card_add}) \implies (\ref{card_iso})$.
  \item $(\ref{card_bot_iff}) \wedge (\ref{card_add}) \wedge (\ref{card_top_finite}) \implies (\ref{card_top_iff_eq})$.
  \end{enumerate}
\end{thm}

Theorem \ref{theorem.card-equivalent-sra} gives structurally simpler alternatives to Kawahara's axioms (\ref{card_univ_comp_meet}) and (\ref{card_univ_meet_comp}), which can be used in Stone relation algebras.
First, (\ref{card_comp_univ}) has only two variables and a simpler right-hand side than (\ref{card_univ_comp_meet}), so it is easier to check if a model satisfies the former.
By Theorem \ref{theorem.card-equivalent-sra}, (\ref{card_comp_univ}) implies (\ref{card_univ_comp_meet}) in the presence of (\ref{card_conv}), and both are equivalent if also (\ref{card_iso}) is assumed.
Second, (\ref{card_univ_meet_conv}) has only two variables and a simpler right-hand side than (\ref{card_univ_meet_comp}), so it too is easier to verify.
Third, (\ref{card_domain_sym}) has only one variable, which makes it structurally simpler than (\ref{card_univ_meet_conv}) or (\ref{card_univ_meet_comp}).
By Theorem \ref{theorem.card-equivalent-sra}, all three are equivalent in the presence of (\ref{card_iso}) and (\ref{card_comp_univ}).

In relation algebras, (\ref{card_iso}) follows from Kawahara's axioms by Theorems \ref{theorem.card-equivalent-sra} and \ref{theorem.card-equivalent-ra}.
Hence we automatically get the above equivalences.
However, in Stone relation algebras it is possible to omit axiom (\ref{card_iso}), which gives a number of alternative choices to (\ref{card_univ_comp_meet}) and (\ref{card_univ_meet_comp}) from the range (\ref{card_comp_univ})--(\ref{card_domain_sym_conv}).
For example, counterexamples generated by Nitpick witness that in the presence of (\ref{card_bot_iff})--(\ref{card_add}),
\begin{itemize}
\item (\ref{card_univ_comp_meet}) and (\ref{card_comp_univ}) are not equivalent, even if (\ref{card_univ_meet_comp}) is assumed additionally;
\item (\ref{card_univ_meet_comp}) and (\ref{card_univ_meet_vector}) are not equivalent, even if (\ref{card_univ_comp_meet}) is assumed additionally;
\item (\ref{card_univ_meet_comp}) and (\ref{card_univ_meet_conv}) are not equivalent, even if (\ref{card_univ_comp_meet}) is assumed additionally.
\end{itemize}
The overall conclusion is that there are many different ways to generalise the cardinality axioms in Stone relation algebras, which cannot be distinguished in relation algebras.

\subsection{Further Axioms for Atoms Below an Element}
\label{section.cardinality.further}

Stone relation algebra $S$ is \emph{atom-rectangular} if every atom in $S$ is a rectangle and \emph{atom-simple} if every atom in $S$ is simple.
It follows that every atomic and atom-simple $S$ is simple.

For example, rectangular atoms are used for reasoning about graphs.
In this context, an edge of a graph can be modelled as a singleton relation, which is a rectangular atom in the full algebra of relations.
The rectangular property allows the contraction of cycles in paths that contain the same edge multiple times \cite{Guttmann2020a}.

The following result gives sufficient conditions for operation $\nAB$ to satisfy axioms other than those covered in Theorem \ref{theorem.nAB-card}.

\begin{thm}
  \label{theorem.nAB-card-sufficient}
  Let $S$ be a Stone relation algebra.
  Then
  \begin{enumerate}
  \item $\nAB$ satisfies (\ref{card_atom_iff}) if $S$ is atomic, atom-rectangular and simple.
  \item $\nAB$ satisfies (\ref{card_atom_iff}) if $S$ is an atomic relation algebra.
  \item $\nAB(\ltop) \leq \nAB(\lone)^2$ if $S$ is atom-rectangular.
  \item $\nAB(\ltop) \geq \nAB(\lone)^2$ if $S$ is atomic and simple with finitely many atoms.
  \item $\nAB$ satisfies (\ref{card_top}) if $S$ is atom-rectangular and atom-simple.
  \item $\nAB$ satisfies (\ref{card_top_finite}) if $S$ has finitely many atoms.
  \end{enumerate}
\end{thm}

\begin{counterexample}
  \label{counterexample.nAB-card-sufficient}
  ~
  \begin{enumerate}
  \item (\ref{card_atom_iff}), (\ref{card_top_iff_eq}) and (\ref{card_top_iff_leq}) do not hold in all atomic and atom-rectangular Stone relation algebras with finitely many atoms.
  \item (\ref{card_atom_iff}), (\ref{card_top_iff_eq}) and (\ref{card_top_iff_leq}) do not hold in all atomic and simple Stone relation algebras with finitely many atoms.
  \item (\ref{card_top_iff_eq}) and (\ref{card_top_iff_leq}) do not hold in all atom-rectangular and atom-simple Stone relation algebras with finitely many atoms.
  \item (\ref{card_top}) does not hold in all atomic relation algebras with finitely many atoms.
  \end{enumerate}
\end{counterexample}

\begin{prf}
  Nitpick found some of the following counterexamples.
  \begin{enumerate}
  \item The Stone relation algebra $S$ of Counterexample \ref{counterexample.non-simple-ideal-point} is atomic and atom-rectangular.
        The only atom in this algebra is $\frac{1}{2}$ but $\nAB(\frac{1}{2}) = \nAB(1) = 1$.
  \item The set $R = \{0,\frac{1}{2},1\} \times \{0,1\}$ of pairs of numbers with the componentwise natural order forms a Stone algebra where $\lbot = (0,0)$ and $\ltop = (1,1)$ and $\cplop$ applies componentwise using $\cplop$ of Stone algebra $S$ of Counterexample \ref{counterexample.non-simple-ideal-point} in the first component and the usual Boolean $\cplop$ in the second component.
        Moreover $R$ forms a Stone relation algebra where $\cnvop$ is identity, $\lone = (0,1)$ and $\cdot$ is defined so that $\lbot$ is its zero, $\lone$ is its unit and the result is $\ltop$ in all other cases.
        Stone relation algebra $R$ is atomic and simple.
        The only atoms in $R$ are $\lone$ and $(\frac{1}{2},0)$, so $\nAB(\cpl{\lone}) = 1$ refutes (\ref{card_atom_iff}) and $\nAB((\frac{1}{2},1)) = \nAB(\ltop) = 2$ refutes (\ref{card_top_iff_eq}) and (\ref{card_top_iff_leq}).
        Note that $R$ does not form a relation algebra because it has $6$ elements, which is not a power of $2$.
  \item The set of integers $\ZZ$ extended by a least element $\lbot$ and a greatest element $\ltop$ forms a Stone relation algebra where $\cnvop$ is identity, $\cdot$ is $\meet$, $\cpl{\lbot} = \ltop$ and $\cpl{x} = \lbot$ for each $x \neq \lbot$.
        This Stone relation algebra has no atoms, and therefore is atom-rectangular and atom-simple.
        Hence $\nAB(x) = 0$ for each $x$.
  \item The set $\{\lbot,\lone,\cpl{\lone},\ltop\}$ of relations over a two-element base set forms an atomic relation algebra which is a subalgebra of the full algebra of relations.
        The atoms in this subalgebra are $\lone$ and $\cpl{\lone}$, so $\nAB(\ltop) = 2$ but $\nAB(\lone) = 1$.
  \QED
  \end{enumerate}
\end{prf}

\subsection{Relation Algebras}
\label{section.relation-algebras}

We now state a converse of Theorem \ref{theorem.nAB-card}.
According to the following result, $\nAB$ is the only cardinality operation satisfying axioms (\ref{card_bot}), (\ref{card_atom}) and (\ref{card_add}) in atomic relation algebras with finitely many atoms.

\begin{thm}
  \label{theorem.card-nAB}
  Let $S$ be an atomic relation algebra with finitely many atoms.
  Let $\#$ be a cardinality operation satisfying axioms (\ref{card_bot}), (\ref{card_atom}) and (\ref{card_add}).
  Then $\# = \nAB$.
\end{thm}

\begin{prf}
  $\# \lbot = 0 = \nAB(\lbot)$ by (\ref{card_bot}).
  For $x \neq \lbot$ we have
  \[
    \# x = \# \bigjoin \AB(x) = \sum \{ \# a \mid a \in \AB(x) \} = \sum \{ 1 \mid a \in \AB(x) \} = \nAB(x)
  \]
  The first equality uses that every atomic relation algebra with finitely many atoms is atomistic, that is, each element is the $\bigjoin$ of the atoms below it.
  The second equality follows by Theorem \ref{theorem.card-equivalent-sra}.\ref{theorem.card-equivalent-sra.card_dist_sup_atoms} using (\ref{card_bot}) and (\ref{card_add}).
  The third equality uses (\ref{card_atom}).
  \QED
\end{prf}

Under the assumptions of Theorem \ref{theorem.card-nAB}, $\#$ satisfies all axioms in Figure \ref{figure.cardinality-axioms} except (\ref{card_top}) by Theorems~\ref{theorem.nAB-card}, \ref{theorem.card-equivalent-sra}, \ref{theorem.card-equivalent-ra} and \ref{theorem.nAB-card-sufficient}.
Nevertheless, the additional axioms including (\ref{card_univ_comp_meet}) and (\ref{card_univ_meet_comp}) are still useful for relation algebras which are not atomic with finitely many atoms.

Theorem \ref{theorem.card-nAB} does not generalise to atomic Stone relation algebras with finitely many atoms: the operation $\#$ on the Stone relation algebra $S$ of Counterexample \ref{counterexample.non-simple-ideal-point} defined by $\# x = 2x$ satisfies (\ref{card_bot})--(\ref{card_top_iff_leq}) and (\ref{card_top_finite}) but $\# 1 = 2 \neq 1 = \nAB(1)$.

The next results give sufficient assumptions about atoms to turn Stone relation algebras into relation algebras.

\begin{thm}
  \label{theorem.atom-sra-ra}
  Let $S$ be an atomic, atom-rectangular and simple Stone relation algebra with finitely many atoms.
  Then $S$ is a relation algebra.
\end{thm}

\begin{prf}
  It suffices to show that every $x \in S$ is \emph{regular}, that is, $\cl{x} = x$.
  This is immediate for $x = \lbot$.
  For $x \neq \lbot$ we first show $x = \bigjoin \AB(x)$; in other words, we show that $S$ is atomistic.
  The inequality $\bigjoin \AB(x) \lleq x$ is immediate.
  The opposite inequality follows from $x \meet \cpl{\bigjoin \AB(x)} = \lbot$ by pseudocomplement properties.
  Assuming the latter is false, there would be an atom $a \lleq x \meet \cpl{\bigjoin \AB(x)}$ since $S$ is atomic.
  But then $a \in \AB(x)$ so $a \lleq \bigjoin \AB(x)$, hence $a \lleq \bigjoin \AB(x) \meet \cpl{\bigjoin \AB(x)} = \lbot$, contradicting that $a$ is an atom.

  Second, we show that $\bigjoin \AB(x)$ is regular by induction over the size of $\AB(x)$ since $S$ has finitely many atoms.
  The induction step follows since $\clop$ distributes over $\join$ in Stone algebras.
  The base cases are singleton sets of atoms, for which we show that every atom $a \in S$ is regular.

  Note that $a \ltop$ is injective since $a \ltop \cnv{(a \ltop)} = (a \ltop \meet \lone) \ltop (a \ltop \meet \lone) \lleq a \ltop \meet \lone \lleq \lone$ using that $S$ is atom-rectangular and $a \ltop \meet \lone$ is an atom by Theorem \ref{theorem.basic}.
  Similarly $\ltop a$ is univalent.
  Moreover $a \ltop$ is surjective and $\ltop a$ is total using that $a$ is simple.
  Hence $a \ltop$ is bijective and $\ltop a$ is a mapping.
  In Stone relation algebras, all bijective elements and mappings are regular and regular elements are closed under $\cdot$ \cite{Guttmann2017b}.
  Therefore $a \ltop a$ is regular.
  But $a = a \ltop a$ since $a$ is a rectangle, which implies that $a$ is regular.
  \QED
\end{prf}

This affects the generalisation of cardinality operations to Stone relation algebras as follows.
By Theorem \ref{theorem.nAB-card}, $\nAB$ satisfies several cardinality axioms in atomic Stone relation algebras.
If we wanted to add (\ref{card_atom_iff}) and (\ref{card_top_finite}) to these, we could use Theorem \ref{theorem.nAB-card-sufficient}.
However, doing so would need the additional assumptions that the Stone relation algebra is simple and atom-rectangular with finitely many atoms.
This would automatically turn it into a relation algebra, which might not be desired.
The same issue would arise if we were to add (\ref{card_top}) and (\ref{card_top_finite}) using Theorem \ref{theorem.nAB-card-sufficient}.

\begin{thm}
  \label{theorem.card-sra-ra}
  Let $S$ be an atom-simple Stone relation algebra with finitely many atoms.
  Assume $\nAB$ satisfies axioms (\ref{card_bot_iff}) and (\ref{card_top}).
  Then $S$ is atomic and atom-rectangular, and hence a representable relation algebra.
\end{thm}

\begin{prf}
  $S$ is atomic since for each $x \neq \lbot$ we have $\nAB(x) \neq 0$ by (\ref{card_bot_iff}), hence there is an atom in $\AB(x)$.

  Next, we consider the mapping $d : S \to S \times S$ defined by $d(x) = (x \ltop \meet \lone,\ltop x \meet \lone)$.
  Recall from Theorem \ref{theorem.basic} that $a \ltop \meet \lone$ and $\ltop a \meet \lone$ are atoms for each atom $a \in S$.
  We prove that $d$ is injective on the set $A$ of atoms of $S$.
  To this end we show that $A$ and the image $d(A)$ have the same finite size.
  In atomic and atom-simple Stone relation algebras, $d(A) = \AB(\lone) \times \AB(\lone)$.
  Hence the number of elements of $d(A)$ is $\nAB(\lone)^2 = \nAB(\ltop)$ by (\ref{card_top}), which is the number of atoms.

  We now show that $S$ is atom-rectangular, which is equivalent to $a \ltop a \lleq \lone$ for all atoms $a \lleq \lone$.
  For this equivalence note that the latter condition implies $a \ltop a = (a \ltop \meet \lone) \ltop (a \ltop \meet \lone) a \lleq a$ since $a \ltop \meet \lone$ is an atom; the converse implication is immediate.

  Hence it suffices to show $a \ltop a \meet \cpl{\lone} = \lbot$.
  Assuming the latter is false, there would be an atom $b \lleq a \ltop a \meet \cpl{\lone}$ since $S$ is atomic.
  It follows that $b \ltop \meet \lone \lleq a \ltop a \ltop \meet \lone \lleq a \ltop \meet \lone$.
  Since both $b \ltop \meet \lone$ and $a \ltop \meet \lone$ are also atoms, $b \ltop \meet \lone = a \ltop \meet \lone$.
  Similarly, $\ltop b \meet \lone = \ltop a \meet \lone$.
  Together we obtain $d(b) = d(a)$, from which injectivity of $d$ gives $b = a$.
  The latter is a contradiction since $a \lleq \lone$ and $b \lleq \cpl{\lone}$.
  \QED
\end{prf}

This has a similar effect to Theorem \ref{theorem.atom-sra-ra}.
If we work with atomic and simple Stone relation algebras that already satisfy (\ref{card_top}), adding (\ref{card_top_finite}) using Theorem \ref{theorem.nAB-card-sufficient} would need the additional assumption of finitely many atoms.
However, this would automatically result in a relation algebra.
Both theorems limit the assumptions we can make if we wish to prove results about Stone relation algebras without an implicit restriction to relation algebras.

Combining Theorems \ref{theorem.nAB-card}, \ref{theorem.card-nAB} and \ref{theorem.card-sra-ra} yields the following result.
It gives sufficient conditions for a cardinality operation to obtain representability of atomic and simple relation algebras with finitely many atoms.

\begin{thm}
  \label{theorem.card-ra}
  Let $S$ be an atomic and simple relation algebra with finitely many atoms.
  Let $\#$ be a cardinality operation satisfying axioms (\ref{card_bot}), (\ref{card_atom}), (\ref{card_add}) and (\ref{card_top}).
  Then $S$ is atom-rectangular and hence representable.
\end{thm}

On the other hand, if we do not assume that the relation algebra is simple, having a cardinality operation does not imply typical sufficient conditions for representability.

\begin{counterexample}
  \label{counterexample.representable}
  There is a representable atomic relation algebra $S$ with finitely many atoms and a cardinality operation satisfying all axioms in Figure \ref{figure.cardinality-axioms}, such that $S$ does not satisfy any of the following conditions:
  \begin{enumerate}
  \item $S$ is atom-rectangular (used in \cite[Theorem 3.2.16]{HenkinMonkTarski1985} for representing cylindric algebras, which model relations of higher arity).
  \item Every atom in $S$ is univalent (used in \cite[Theorem 4.29]{JonssonTarski1952} for representing relation algebras).
  \item For every $x \in S$ with $\lbot \neq x \lleq \lone$, there is a $y \in S$ with $\lbot \neq y \lleq x$ and $y \ltop y \lleq \lone$ (used in \cite[Theorem (C)]{Maddux1991} for representing relation algebras).
  \end{enumerate}
\end{counterexample}

\begin{prf}
  Let $R$ be the four-element relation algebra of Counterexample \ref{counterexample.atom_univalent}.
  Then $R$ is atomic with atoms $\lone$ and $\cpl{\lone}$, so $\nAB(\ltop) = 2$ and $\nAB(\lone) = \nAB(\cpl{\lone}) = 1$ and $\nAB(\lbot) = 0$.
  By Theorems \ref{theorem.nAB-card}, \ref{theorem.card-equivalent-sra}, \ref{theorem.card-equivalent-ra} and \ref{theorem.nAB-card-sufficient}, $\nAB$ satisfies all axioms in Figure \ref{figure.cardinality-axioms} except (\ref{card_top}).

  To obtain (\ref{card_top}) we consider $S = R \times R$.
  The Cartesian product of two relation algebras is a relation algebra where operations are applied componentwise.
  It follows that $S$ can be represented as a subalgebra of the full algebra of relations over a four-element base set.
  Likewise, $S$ is atomic with atoms $(x,\lbot)$ and $(\lbot,y)$ using atoms $x, y \in R$.
  Using the operation $\# (x,y) = \# x + \# y$ for $x, y \in R$, the Cartesian product construction preserves cardinality axioms (\ref{card_bot_iff}), (\ref{card_atom_iff}), (\ref{card_conv}), (\ref{card_add}), (\ref{card_univ_comp_meet}) and (\ref{card_univ_meet_comp}).
  By Theorems \ref{theorem.card-equivalent-sra}, \ref{theorem.card-equivalent-ra} and \ref{theorem.nAB-card-sufficient}, $\#$ satisfies all axioms in Figure \ref{figure.cardinality-axioms} except (\ref{card_top}).
  Moreover $\# (\ltop,\ltop) = \# \ltop + \# \ltop = \nAB(\ltop) + \nAB(\ltop) = 2 + 2 = 4$ and $\# (\lone,\lone) = \# \lone + \# \lone = \nAB(\lone) + \nAB(\lone) = 1 + 1 = 2$, hence also (\ref{card_top}) holds in $S$.

  Nitpick found the following counterexamples in $S$.
  \begin{enumerate}
  \item $(\lone,\lbot)$ is an atom but not rectangular as $(\lone,\lbot) (\ltop,\ltop) (\lone,\lbot) = (\lone \ltop \lone,\lbot \ltop \lbot) = (\ltop,\lbot) \not\lleq (\lone,\lbot)$.
  \item $(\lbot,\cpl{\lone})$ is an atom but not univalent since $\cnv{(\lbot,\cpl{\lone})} (\lbot,\cpl{\lone}) = (\cnv{\lbot},\cpl{\lone}\cnvop) (\lbot,\cpl{\lone}) = (\lbot,\cpl{\lone}) (\lbot,\cpl{\lone}) = (\lbot \lbot,\cpl{\lone} \, \cpl{\lone}) = (\lbot,\ltop) \not\lleq (\lone,\lone)$.
  \item Let $x = (\lone,\lone)$.
        Then $(\lbot,\lbot) \neq x \lleq (\lone,\lone)$.
        Assume $(\lbot,\lbot) \neq y \lleq x$.
        Then $y = (\lbot,\lone)$ or $y = (\lone,\lbot)$ or $y = x$.
        First, $(\lbot,\lone) (\ltop,\ltop) (\lbot,\lone) = (\lbot \ltop \lbot,\lone \ltop,\lone) = (\lbot,\ltop) \not\lleq (\lone,\lone)$.
        Second, $(\lone,\lbot) (\ltop,\ltop) (\lone,\lbot) = (\lone \ltop \lone,\lbot \ltop \lbot) = (\ltop,\lbot) \not\lleq (\lone,\lone)$.
        Third, $(\lone,\lone) (\ltop,\ltop) (\lone,\lone) = (\lone \ltop \lone,\lone \ltop \lone) = (\ltop,\ltop) \not\lleq (\lone,\lone)$.
  \end{enumerate}
  In fact, condition 1 implies condition 3 for any atomic Stone relation algebra $S'$:
  if $x \in S'$ with $\lbot \neq x \lleq \lone$, there is an atom $a \in S'$ with $a \lleq x$, hence $a \ltop a \lleq a \lleq x \lleq \lone$ by condition 1.
  \QED
\end{prf}

\section{Open Problem and Future Work}
\label{section.open}

A referee has raised the question whether every point is an ideal-point if there exists an ideal-point.
Theorem \ref{theorem.point-axiom} gives an affirmative answer if we assume the point axiom.
We intend to further look into the relationship between points and ideal-points with a view to answering this question without the point axiom.
We also plan to study the consequences of using points instead of ideal-points in the point axiom.

In this paper we restrict $\bigjoin P$ to finite non-empty sets $P$.
This restriction is inherited from a library used by the formally verified theories in Isabelle/HOL.
Many results in this paper can be extended to include $\bigjoin$ of the empty set and some results may generalise to $\bigjoin$ of arbitrary sets.
Formally verifying such extensions requires a change or modification of the library, so we defer these to future work.
We would also like to study applications of the cardinality axioms to graph problems following \cite{BerghammerDanilenkoHoefnerStucke2016}.

\paragraph{Acknowledgement}
Hitoshi Furusawa and Walter Guttmann thank the Japan Society for the Promotion of Science for supporting part of this research through a JSPS Invitational Fellowship for Research in Japan.
We thank the anonymous referees for their helpful feedback.

\bibliographystyle{fundam}
\bibliography{cardinality}

\end{document}